\documentclass{pasj00}
\onecolumn
\RelativeFigurePath{figs}
\usepackage{color}

\usepackage{lscape}

\begin{document}
\SetRunningHead{K. Hayasaki, H. Saito and S. Mineshige}
               {Binary Black Hole Accretion Flows From a Misaligned Circumbinary Disk}
\Received{2012/??/??}% (in the 1st incomplete version)
\Accepted{2012/??/??}%{yyyy/mm/dd}

\title{Binary Black Hole Accretion Flows From a Misaligned Circumbinary Disk}
%in a Merged Galactic Nuclei}

%%% end:list of authors

%%% Please use the following style in case that sorting by 
%%% affilation is impossible. 
%
 \author{
   Kimitake \textsc{Hayasaki}\altaffilmark{1,2},
   Hideki \textsc{Saito}\altaffilmark{2,3},
   and
   Shin \textsc{Mineshige}\altaffilmark{2}}
 \altaffiltext{1}{Korea Astronomy and Space Science Institute, Daedeokdaero 776, Yuseong, Daejeon 305-348, Korea}
 \altaffiltext{2}{Department of Astronomy, Kyoto University, Sakyo-ku, Kyoto 606-8502, Japan}
% \email{eeeee@xxx.xxx.xx.xx}
 \altaffiltext{3}{Nagano City Museum, Oshimadamachi, Nagano, Nagano 381-2212, Japan}
  \email{kimi@kasi.re.kr; kimi@kusastro.kyoto-u.ac.jp}

%% `\KeyWords{}' always has to be placed before `\maketitle'.
%Do NOT move this preamble from here!

\maketitle

%%%%%%% Abstract %%%%%%%%

\begin{abstract}
We study the basic properties of accretion flows onto 
binary supermassive black holes, including the cases 
in which a circumbinary disk is misaligned with the 
binary orbital plane, by means of three-dimensional 
Smoothed Particle Hydrodynamics simulations. We find 
that a circular binary system with a misaligned circumbinary 
disk normally produces a double peaked mass-accretion-rate 
variation per binary orbit. This is because each black hole passes 
across the circumbinary disk plane and captures gas twice 
in one orbital period. Even in misaligned systems, however, 
a single peaked mass-accretion-rate variation per binary orbit is produced, 
if the orbital eccentricity is moderately large ($e\gtrsim0.3$).
The number of peaks in mass accretion rates can be understood 
simply in terms of the orbital phase dependence of the distance 
between each binary black hole and its closest inner edge 
of the circumbinary disk. In the cases of eccentric binary black 
holes having different masses, the less massive black hole can 
get closer to the circumbinary disk than the massive one, thus 
tidally splitting gas from its inner edge, but the created gas flows 
are comparably captured by both black holes with a short time 
delay. As a consequence, the combined light curve shows 
periodic occurrence of double-peaked flares with a short interval. 
This may account for the observed light variations of OJ287.
\end{abstract}

%
%%%%%%%%%%%
\section{Introduction}
%%%%%%%%%%%
%
% 1st paragraph
Hierarchical structure formation scenario tells us that 
a galaxy grows by merger of smaller galaxies. 
The recently confirmed correlation between the mass of 
supermassive black holes (SMBHs) and the mass or 
luminosities of the bulge of their host galaxies 
strongly support the idea that SMBHs have grown with 
the growth of their host galaxies \citep{mag98,geb00,fm00}. 
These relationships suggest that each SMBH in the center of each galaxy
should have evolved toward coalescence in a merged galaxy. 
If this is the case, a binary of SMBHs 
should be formed in a merged galactic nucleus before 
two black holes finally coalesce, 
yet no binary SMBHs have clearly been identified so far 
except for some candidates at large separations ($\sim\rm{kpc}$) (e.g., \cite{ko03}).

%
% 2nd paragraph
The merger of two SMBHs processes via three stages \citep{bege80}:
First, each black hole sinks into a common center of a merged galactic nucleus 
by the dynamical friction with the surrounding field stars and gas \citep{es05,dotti07}. 
When the separation between the two black holes becomes as short as one parsec or so, 
the dynamical friction is no longer efficient and a hard binary is formed \citep{may07}.
The separation of such a hard binary should have been reduced by some unknown mechanism.
When the separation becomes as short as one parsec or less, finally,
the binary rapidly merge by emitting gravitational wave radiation to become a single SMBH.
However, there has been many discussions about by what mechanism the binary orbit decays 
in the second stage. One of predominant candidates is the interaction between the binary 
and the circumbinary disk \citep{iv99,gr00,an02,kh09,hai09,cu09,lo09,hui10,nix11a,kocs12a,kocs12b}.

% 3rd paragraph
It is an observational challenge how to identify 
binary SMBHs on such a subparsec scale (see \citet{ko06} for a review).
Several ways have been proposed:
Periodic optical and radio outbursts (e.g., OJ 287) \citep{sill88,val11}, 
wiggled patterns of the radio jet, indicating precessional motions 
on a parsec scale \citep{yo85,lr05}, X-shaped morphology of radio 
lobes \citep{me02}, double-peaked broad emission lines in active galactic nuclei 
(AGNs) \citep{ga96} and distant quasars \citep{bo09,dotti09,bl09,mont11,mont12,era12}, 
double compact cores with a flat radio spectrum \citep{ro06}, 
orbital motion of the compact core with a periodic flux variation 
\citep{su03,ig10}, and so on. 

% 4th paragraph
With successive discoveries of binary black hole candidates 
and ongoing constructions of advanced gravitational wave 
detectors such as ${\it eLISA}$\citep{as12}, much attention 
has been recently paid to electromagnetic signatures from 
binary SMBH systems, such as afterglows \citep{mp05,tk10,tk12}, 
precursors \citep{cp09,bode10,kh11,fa11,fa12,bode12}, periodic 
emissions \citep{al96,lv96,hms07,hmh08,bo08,mm08,rdscc11, 
sesa12,shi12,noble12,dza12}, and a dual-jet structure \citep{pal10,mos10},
in the context of a massive black hole coalescence.
There are also several theoretical studies on the observational 
methodology to probe the presence of a gaseous disk around 
coalescing binary black holes through the waveform analyses of 
the emitted gravitational waves \citep{kocs11,yunes11,hytm12}.

% 5th paragraph
In most of the previous studies, it has been assumed that 
the circumbinary disk is aligned with the binary orbital plane. 
However, the angular momentum vector of the binary does 
not always coincide with that of the circumbinary disk because 
the circumbinary disk would be formed independently of the 
formation of binary SMBHs. Therefore, the orientation of a 
circumbinary disk plane can be taken arbitrarily with respect 
to the binary orbital plane. It has been shown quite recently 
that the circumbinary disk which rotates in such an opposite 
direction as the binary rotation can be stable, if the misalignment 
angle between the binary orbital plane and the initial circumbinary 
disk plane is more than $\pi/2$ \citep{nix12}, whereas the 
circumbinary disk which rotates in a prograde direction for the 
binary is stable, if the misalignment angle is less than $\pi/2$. 
However, accretion flows onto binary SMBHs around the 
misaligned circumbinary disk rotating in a prograde direction 
have been poorly investigated. The misalignment angle is 
expected to produce complex light variations which never 
appear otherwise. Especially, two periodic outbursts per orbital 
period may be obtained, since there are two epochs in one orbital 
period, when the distance between the binary and the inner edge 
of circumbinary disk is shortest. By contrast, the binary orbital 
eccentricity produces a single, steep peak per binary orbit in a 
coplanar system \citep{hms07,rdscc11}.

% final paragraph
In this paper, we focus our study on basic properties of accretion 
flows onto binary SMBHs including the cases of a misaligned 
circumbinary disk. In the next section, we will describe our models 
and methods of calculations. The numerical results will be presented 
in section~3. In section~4, we provide a simple model in order to 
understand the numerical results. The final section is devoted to 
summary and discussion.

%
%%%%%%%
% Figure 1
%%%%%%%
%
\begin{figure}
\begin{center}
\includegraphics[width=12cm]{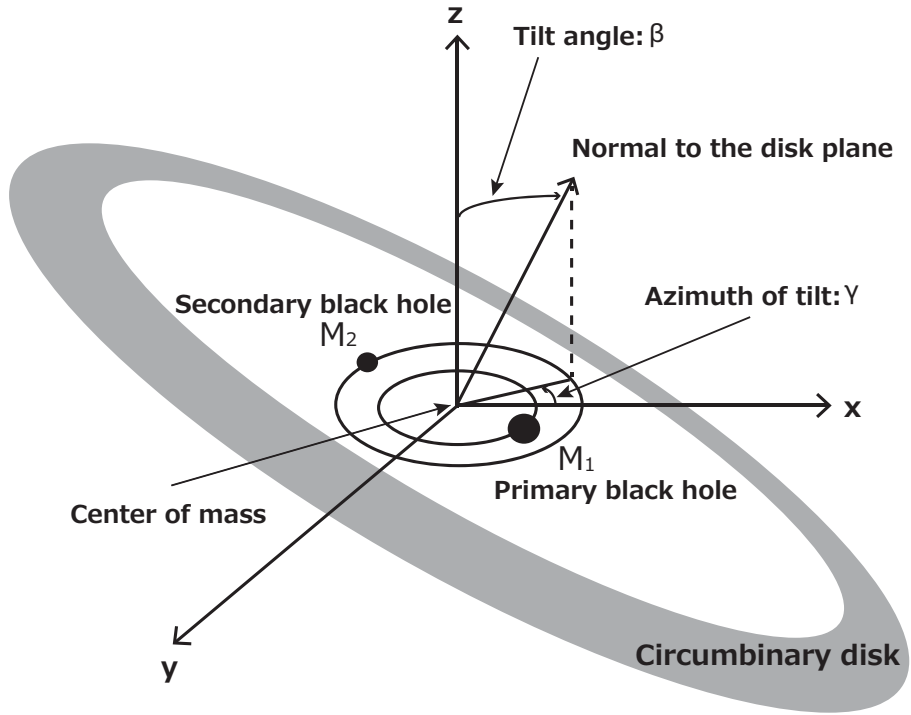}
\end{center}
\caption{
Configuration of our model.There are two angles ($\beta, \gamma$) 
which specify the orientation of the circumbinary disk plane with respect to 
the binary orbital plane ($x$-$y$ plane).
}
\label{fig:schmatic}
\end{figure}
%%%%%%
%
%
%%%%%%%%%%%%%%%%%%%%%%%
\section{Our models and calculation methods}
\label{sec:num-pro}
% Section 2
%%%%%%%%%%%%%%%%%%%%%%%
%
In this section, we first explain our models 
and next describe our calculation methods.
%
%%%%%%%%%%%%%%%%
\subsection{Initial settings}
%%%%%%%%%%%%%%%%
%

% 1st paragraph
Figure~\ref{fig:schmatic} illustrates a schematic picture of the setting of our model; 
binary black holes rotating each other are surrounded by
a misaligned circumbinary disk.
The binary is put on the $x$-$y$ plane with its center of mass being at the origin. 
The masses of the primary and secondary black holes are represented 
by $M_1$ and $M_2$, respectively. We put a circumbinary disk around the origin.
The unit vector of specific angular momentum of the circumbinary disk is expressed 
by (e.g. \cite{pr96})
\begin{equation}
\mbox{\boldmath $j$}_{\rm{d}}=(\cos\gamma\sin\beta,\sin\gamma\sin\beta,\cos\beta),
\label{eq:damvec}
\end{equation}
where $\beta$ is the tilt angle between the circumbinary disk plane and the 
binary orbital plane, and $\gamma$ is the azimuth of tilt. 
%(i.e. $\gamma+\pi/2$ represents the azimuth of the ascending node).
The position vector of each black hole is given by
\begin{equation}
\mbox{\boldmath $r$}_{i}=(r_{i}\cos\phi, r_{i}\sin\phi, 0) \hspace{2mm}[i=1,2],
\label{eq:ri}
\end{equation}
where
\begin{equation}
r_{i}=\eta_{i}\frac{a(1-e^2)}{1+e\cos\phi} 
\label{eq:rphi}
\end{equation}
and $\phi$ is the true anomaly (hereafter, we regard the true anomaly as the orbital phase) 
(e.g., \cite{mcd99}), $\eta_1\equiv q/(1+q)$ and $\eta_2\equiv 1/(1+q)$ with binary mass 
ratio $q=M_2/M_1$, and $a=a_1 + a_2$ is the semi-major axis of the binary, where 
$a_1\equiv\eta_1a$ and $a_2\equiv\eta_2a$.

% 2nd paragraph
The circumbinary disk initially has a radially uniform density profile
between the initial radius of the disk-inner edge, $r_{\rm ini}$, 
and the radius of $r = r_{\rm ini} + 0.05a$ (i.e., the initial width of the circumbinary disk is set to be $0.05a$).
The material in the circumbinary disk rotates around the origin on the circular orbit 
with the Keplerian rotation velocity. The vertical density structure of the circumbinary disk is exponential;
i.e., we assume a hydrostatic balance with constant temperature in the vertical direction. 
The initial mass of the circumbinary disk is $1.0\times 10^{-4}M_\odot$. The disk temperature is assumed
to be $T=30700{\rm K}$ everywhere. Note that this corresponds to the
typical central temperature of a standard disk at $r=2a$ around a single black hole with
$10^{8}M_{\odot}$ for a given mass input rate of 
$\dot{M}_{\rm inj}=1.0M_{\odot}~{\rm yr}^{-1}$ \citep{kato08}. 

%
% 3rd paragraph
In Table \ref{tb:t1}, we summarize common model parameters 
for all models in the present study:
total black hole mass ($M_{\rm{1}}+M_{\rm{2}}$), 
semi-major axis of the binary orbit ($a$), 
initial mass of the circumbinary disk ($M_{\rm disk}$), 
mass injection rate ($\dot{M}_{\rm inj}$), 
disk temperature $T_{\rm d}$, and Shakura-Sunyaev viscosity 
parameter $\alpha_{\rm{SS}}$ \citep{ss73}. The relation between 
$\alpha_{\rm{SS}}$ and the SPH artificial viscosity parameters will 
be described in section~\ref{subsec:method}.
The binary orbital period, $P_{\rm{orb}}$, is estimated to be 
approximately $9.4\,\rm{yr}$ by using the Keplerian third law.

%
%%%%%%%%%%%
% Table 1
%%%%%%%%%%%
%
\begin{table}[!h]
  \caption{Common model parameters}
   \begin{center}
     \begin{tabular}{ll}
       \hline \hline
       Total black hole mass  & $M_{\rm 1}$+$M_{\rm 2}$=$10^{8}M_{\odot}$  \\
       Semi-major axis & $a=0.01~{\rm pc}$ \\ 
       (Orbital period) & ($P_{\rm orb}=2\pi a^{3/2}/\sqrt{G(M_1+M_2)}\approx9.4~{\rm yr}$)  \\
       Initial disk mass &  $M_{\rm disk} =1.0\times 10^{-4}M_\odot$ \\
       mass injection rate & $\dot{M}_{\rm inj}=1.0M_{\odot}~{\rm yr}^{-1}$ \\
       Disk temperature & $T_{\rm{d}}=30700~{\rm K}$ \\ 
       Disk viscosity & $\alpha_{\rm{SS}}=0.1$ \\
       \hline
     \end{tabular}
   \label{tb:t1}
%   \tablecomments{{\bf More detailed explanation can be found on the text.}}
   \end{center}
\end{table}
%%%%%%%%%%%
%
%%%%%%%%%%%%%%%%
\subsection{Boundary conditions}
\label{sec:ibcond}
%%%%%%%%%%%%%%%%
%

% 1st paragraph
Gas particles are added to the outer edge of the circumbinary disk 
from its outside in an arbitrary angles at a constant rate of 
$\dot{M}_{\rm inj}=1.0M_{\odot}~{\rm yr}^{-1}$.
The inner edge of the circumbinary disk is determined 
by the balance between the tidal/resonant torque exerted 
by the binary black holes and the viscous torque of the circumbinary disk.
We take an initial inner edge radius of the circumbinary disk to be 
$r_{\rm{ini}}=2.5$ for $e=0.5$. In an equal-mass and circular binary, 
on the other hand, we take an initial inner edge radius as $r_{\rm{ini}}=1.68a$, 
corresponding to the tidal truncation radius where the tidal torque of the 
binary equals to the viscous torque of the circumbinary disk (\cite{pap77}: 
see also Table $1$ of \cite{al94}). 

% 2nd paragraph
We set the outer calculation boundary at $r=6.0a$, which is sufficiently far from
the disk region so that the outer boundary should not affect the flow dynamics 
in the binary SMBH system. The SPH particles passing outward across the 
outer calculation boundaries are removed from the simulation box. 

% 3rd paragraph
The accretion radius depends on the mass of each black hole.
The black holes are modeled by sink particles with the fixed accretion radius 
of $r_{\rm acc}=0.1a$ or $0.05a$, depending on the black hole mass.
Note that each accretion radius is two orders of magnitude larger than 
the Schwarzshild radius of each black hole. Numerically, we remove all the 
particles that enter the region inside the accretion radius.

%
%%%%%%%%%%%%%%%%
\subsection{Numerical method}
\label{subsec:method}
%%%%%%%%%%%%%%%%
%

% 1st paragraph
The simulations presented below are performed 
with the three-dimensional (3D) SPH code, 
which is based on a version originally developed 
by \cite{benz90a,benz90b,bate95} 
and has been extensively used for various systems by many authors 
(e.g., \cite{ato02}; \cite{hms07}; \cite{ato10}; \cite{tak12}).
The SPH equations are composed of a mass 
conservation equation, a momentum equation 
with the SPH standard artificial viscosity, and an 
isothermal equation of state in substitution for 
an energy equation. These equations with the standard 
cubic-spline kernel are integrated using a second-order 
Runge-Kutta-Fehlberg integrator with individual time steps 
for each particle and a variable smoothing length \citep{bate95}, 
which results in saving enormous computational time 
when a large range of dynamical timescales are involved.

% 2nd paragraph
The artificial viscosity commonly used in SPH consists 
of two terms: a term that is linear in the velocity differences 
between particles, which produces a shear and bulk viscosity, 
and a term that is quadratic in the velocity differences, which is 
needed to eliminate particle interpenetration in high Mach number 
shocks. The parameters $\alpha_{\rm SPH}$ and $\beta_{\rm SPH}$ 
control the linear and quadratic terms, respectively. In the simulations 
shown in this paper, the artificial viscosity is adjusted so as to keep 
the Shakura-Sunyaev viscosity parameter $\alpha_{\rm SS}=0.1$ \citep{ss73}, 
using the approximate relation $\alpha_{\rm SS} = 0.1\, \alpha_{\rm SPH}\, 
h/H$ and $\beta_{\rm SPH}=0$ (see Section~2.2 of \cite{hms07}), 
where $h$ and $H$ are the smoothing length of individual particles 
and the scale-height of the circumbinary disk, respectively.

% 3rd paragraph
Sink particles (i.e. black holes) are orbiting around each other, 
following the Keplerian third law, because perturbations 
by the interaction with SPH particles are negligible. 
In all models described in next subsection, total run time is $60$ in a unit of $P_{\rm orb}$, 
and the simulation time $t$ is normalized by $P_{\rm{orb}}$ through this paper.
The orbital phase is forwardly shifted by $0.03$ at the end of run ($t=60$) 
to correct the accumulation of numerical errors.

%
%%%%%%%%%%%
% Table 2
%%%%%%%%%%%
%
 \begin{table}[!h]
  \caption{Individual model parameters.}
   \begin{center}
     \begin{tabular}{ccccccccccc}
       \hline \hline
         Model & $q\equiv M_{\rm 2}/M_{\rm 1}$ & $e$ & $r_{\rm in}/a$ & ($\beta, \gamma$) \\
       \hline \hline
       A1 & 1.0 & 0.0 & 1.68 & $(0, 0)$ \\
       A2 & 1.0 & 0.0 & 1.68 & $(\pi/6, \pi/2)$ \\
       A3 & 1.0 & 0.0 & 1.68 & $(\pi/4, \pi/2)$ \\
       A4 & 1.0 & 0.0 & 1.68 & $(\pi/6, 0)$       \\
       \hline
       B1 & 1.0 & 0.5 & 2.50 & $(0, 0)$  \\
       B2 & 1.0 & 0.5 & 2.50 & $(\pi/6, \pi/2)$ \\
       B3 & 1.0 & 0.5 & 2.50 & $(\pi/4,\pi/2)$ \\
       B4 & 1.0 & 0.5 & 2.50 & $(\pi/6,0)$ \\
       \hline
       C1 & 0.5 & 0.5 & 2.50 & $(0,0)$ \\  
       C2 & 0.5 & 0.5 & 2.50 & $(\pi/6,\pi/2)$ \\
       C3 & 0.5 & 0.5 & 2.50 & $(\pi/4,\pi/2)$ \\
       \hline
     \end{tabular}
   \label{tb:t2}
   \end{center}
 \end{table}
%%%%%%

%
%%%%%%%%%%%%%%%%
\subsection{Calculated models}
\label{subsec:models}
%%%%%%%%%%%%%%%%
%

% 1st paragraph
In the present study we are concerned with 
the observable quantities for various configurations of 
binary SMBHs with the circumbinary disk.
Accordingly, we calculated eleven models, in total, by 
varying the binary mass ratio, binary orbital eccentricity, 
title angle, and azimuth of tilt. In Table \ref{tb:t2}, we 
summarize the model parameters: from the left to right, 
model number (first column), mass ratio of the secondary 
black hole to the primary one (second column), orbital 
eccentricity (third column), initial radius of the inner edge 
of the circumbinary disk in units of $a$ (fourth column), 
and tilt angle and azimuth of tilt $(\beta, \gamma)$ (fifth column, 
see Figure~\ref{fig:schmatic} and equation~(\ref{eq:damvec}) 
for definitions of $\beta$ and $\gamma$).

% 2nd paragraph
While Model A is an equal-mass and circular binary,
Models B and C are eccentric binaries with equal black hole masses (Models B) and 
unequal black hole masses (Model C). Through this paper, we assign Model A2 as a fiducial model.

%
%%%%%%%%%%%%%%%%%%%%%%%%%%%%%%%%%%%%%%%%
\section{Accretion flows from a misaligned circumbinary disk to binary black holes}
\label{sec:3}
%%%%%%%%%%%%%%%%%%%%%%%%%%%%%%%%%%%%%%%%
%

In this section, we examine how the basic properties of accretion flows onto each black hole 
depend on $\beta$, $\gamma$, $e$, and $q$, by performing 3D SPH simulations.

%
%% Figure 2 %%
%
\begin{figure}
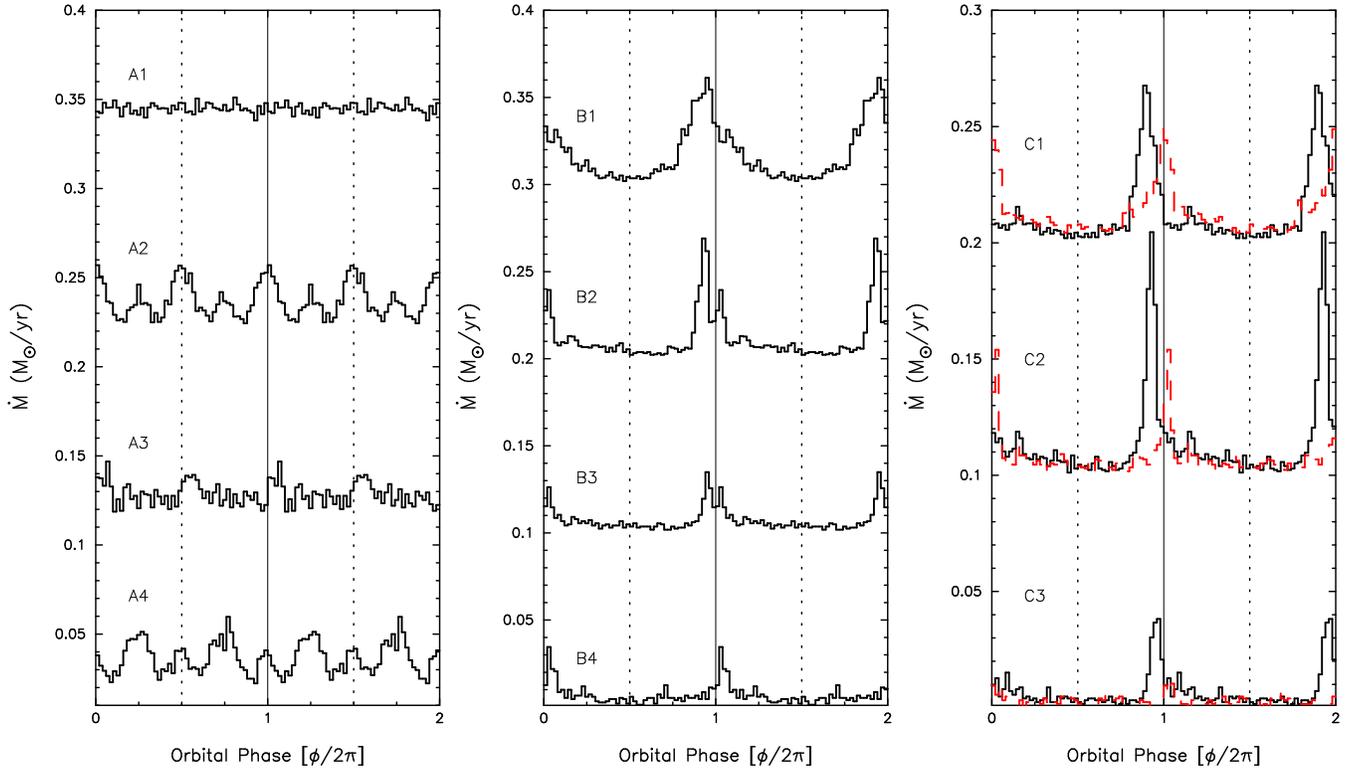

\begin{center}
\includegraphics[angle=0,scale=.50]{figure2a.ps}
\includegraphics[angle=0,scale=.50]{figure2b.ps}
\includegraphics[angle=0,scale=.50]{figure2c.ps}
\caption{
Orbital phase dependence of mass accretion rates for all models.
Models A1--A4,  Models B1--B4, and Models C1--C3 are shown 
in the left panel, the middle panel, and the right panel, respectively.
The data is folded on the orbital period over $40 \leq t \leq 60$.
Here, we redefine the orbital phase in order that the phase 0 corresponds 
to the epoch of the periastron passage (see section 2.3).
The binary black holes are at the periastron (or at the apastron) 
at phase 0.0 (0.5) for eccentric binaries (Models B1--C3).
For clarity, we vertically offset the mass accretion rates for Model A1 and B1 by +0.3, 
for Models A2, B2, and C1 by +0.2, and for Models A3, B3, and C2 by +0.1 
with respect to those of Models A4, B4, and C3, respectively.
In the right panel, the solid line and dashed (red) line represent the mass 
accretion rates onto the primary black hole and that of the secondary one, respectively.
}
\label{fig:mdotL}
\end{center}
\end{figure}
%%%%%%
%%%%%
% Table 3
%%%%%
\begin{table}[!h]
  \caption{
Summary of averaged mass accretion rates, averaged circularization radii, 
and the number of SPH particles at the end of the run (t=60) for all models. 
Since the mass is injected to the circumbinary disk at the rate of $1M_\odot/\rm{yr}^{-1}$, 
the averaged mass accretion rates indicate the fraction of the mass accretion 
rate onto each black hole to the mass injection rate.
  }
   \begin{center}
     \begin{tabular}{cccccccccccc}
       \hline \hline
       Model  
@   & $\langle{\dot{M}_{1}}\rangle$ & $\langle{\dot{M}_{2}}\rangle$  
       & $\langle{r_{\rm{c},1}}\rangle$ & $\langle{r_{\rm{c},2}}\rangle$ & $N_{\rm{SPH}}$\\
       & [$M_{\odot}{\rm yr^{-1}}$] & [$M_{\odot}{\rm yr^{-1}}$] & [a] & [a] \\
       \hline
       \hline
       A1 & $0.045$ & $0.045$ & $0.033$ & $0.033$ & 62157 \\ 
       A2 & $0.037$ & $0.036$ & $0.062$ & $0.062$ & 67774 \\ 
       A3 & $0.028$ & $0.028$ & $0.055$ & $0.056$ & 53866 \\ 
       A4 & $0.037$ & $0.037$ & $0.063$ & $0.063$ & 81149 \\ 
       \hline
       B1 & $0.018$ & $0.017$ & $0.038$ & $0.034$ & 61022 \\
       B2 & $0.012$ & $0.012$ & $0.059$ & $0.058$ & 90168 \\
       B3 & $0.007$ & $0.007$ & $0.051$ & $0.051$ & 55987 \\
       B4 & $0.007$ & $0.007$ & $0.053$ & $0.057$ & 58000 \\
       \hline
       C1 & $0.013$ & $0.013$ & $0.037$ & $0.017$ & 74489 \\  
       C2 & $0.014$ & $0.008$ & $0.056$ & $0.028$ & 54729 \\ 
       C3 & $0.007$ & $0.003$ & $0.056$ & $0.025$ & 51317 \\ 
       \hline
     \end{tabular}
   \label{tb:t3}
   \end{center}
\end{table}
%%%%%%%%%

%
%%%%%%%%%%%%%%%%%%%%%%%%%%%%
\subsection{Mass accretion rates onto binary black holes}
\label{sec:mdotL}
%%%%%%%%%%%%%%%%%%%%%%%%%%%%
%
% 1st paragraph
We first show in Figure \ref{fig:mdotL} the orbital phase dependence 
of mass accretion rates for all models.
Through this paper, the mass accretion rate is calculated by counting 
the number of SPH particles entering the accretion radius of each black hole 
per unit time. Each mass accretion rate is folded on the orbital period over 
$40\le{t}\le60$. For eccentric binaries (Models B1--C3), the binary is at the 
periastron (or at the apastron) at phase 0.0 (0.5). For clarity, we vertically offset 
the mass accretion rates for Model A1 and B1 by +0.3, for Models A2, B2, 
and C1 by +0.2, and for Models A3, B3, and C2 by +0.1 with respect to those 
of Models A4, B4, and C3, respectively.

%
% 2nd paragraph
\citet{hms07} already studied the aligned disk case (i.e., $\beta=0$) 
corresponding to Model~A1. Since the mass continuously falls onto the 
binary from the circumbinary disk, there is no remarkable orbital-phase 
dependence in a circular orbit. As a result, no periodic variations are 
produced in neither mass accretion rates onto the two black holes nor 
the luminosity of accretion disks surrounding the two black holes. 
In other words, the accretion proceeds in a quasi-steady fashion, and, 
hence, no large light variations but small amplitude, random fluctuations 
are observed. More details are seen in Figures 2, 4 and 8 of \citet{hms07}. 
We should note, however, that even in the case of a circular binary, periodic 
variations may be produced by the formation of spiral patterns excited on 
the circumbinary disk \citep{mm08,dza12}. Such features never appear 
in our calculations with limited spatial dimension of the circumbinary disk 
and, so, will be explored in future.

% 3rd paragraph
The situations differ markedly, if the circumbinary disk is misaligned with 
respect to the binary orbital plane. In Models A2 and A3, the circumbinary 
disks are tilted by $\beta = \pi/6$ and $\pi/4$, respectively, while $\gamma$ 
is kept the same (i.e., $\gamma=\pi/2$). Both models produce mass-accretion-rate 
variations with (more than) two peaks in one orbital period, as shown in the 
left panel of Figure~\ref{fig:mdotL}. This can be understood in terms of the 
geometrical effect. In these misaligned systems, two black holes stay away 
from the circumbinary disk plane in most of time, but twice per orbital period 
they cross the circumbinary disk plane. Thus, there are two chances in one 
orbital period that black holes approach the inner region of the circumbinary 
disk and strip gas from the circumbinary disk. This effect makes the mass 
accretion rates enhanced around the phases of $0.0$ and $\pi$. In addition, 
there are two minor peaks in the mass accretion rate of Model A2. 
Their origin is, however, unclear at this moment.

% 4th paragraph
We also examine how the mass-accretion-rate variations depend on the azimuth 
of tilt ($\gamma$) in Model~A4, in which we assign $\gamma=0$, keeping 
other parameters the same as those in Model A2. The resultant variation patterns 
of the mass accretion rate is the same as those in Model A2 but with a horizontal 
shift by $\pi/4$, as is expected.

% 5th paragraph
In Models B1--B4, more clear-cut periodic light variations in mass accretion rates 
are obtained as a natural consequence of the eccentric binary orbit with $e=0.5$. 
The coplanar case (i.e., Model B1) was also studied by \citet{hms07}, 
who found that binary SMBHs periodically emit intense pulses as a result 
of the periodic interaction between the binary and the circumbinary disk.
The pulse period is therefore the orbital period.
The variation amplitude is estimated to be a factor of $6-7$ for Models B1 and B2, 
while it is a factor of $3-4$ for Model B3 and B4.
Obviously, interactions are strongest when the black holes are at the apastron, 
where the black holes are closest to the inner edge of the circumbinary disk.
However, the mass accretion rate reaches a peak not at the phase $\phi \sim \pi$ 
(near apastron passage) but at $\phi\sim 0.0$ (near periastron passage).
This is because it takes about half an orbital period
for the gas stripped from the circumbinary disk to freely fall onto the black holes.
In Model~B4,  we assign $\gamma=0$, keeping other parameters 
the same as those in Model B2.
The resultant mass accretion rate has a single peak per binary orbit, 
and the variation amplitude is smaller than that of Model B2.
It is interesting to note that double-peak nature, which was observed 
in Models A2 and A3, is no longer noticeable in Models B2--B4. 
This is because the effect of orbital eccentricity is stronger than 
both the misalignment effect and the rotation effect in the azimuth direction, 
thus erasing the double-peak nature (see section~\ref{sec:sam}).
Note that it seems that the mass-accretion-rate variations have a double peaked 
structure at $\phi\sim0$ in Model B2 and B3, but such a structure may be numerical 
artifacts.

% 6th paragraph
We finally performed simulations of Models C1--C3 to see what variations in mass 
accretion rates are produced by the binary with different masses. The results are 
shown for Models C1--C3 in the right panel, where the solid line and dashed (red) 
line represent the mass accretion rates onto the primary and secondary black holes, 
respectively. Single peaked light variations are obtained for these models. The mass 
accretion rate onto the secondary black hole is slightly smaller than those onto the 
primary black hole. Remarkably, the peak phases slightly differ among the two black 
holes. This is understood as follows. The gas in the inner edge of the circumbinary 
disk is tidally stripped at the apastron when the secondary black hole gets closest to 
the circumbinary disk. The gas freely infalls onto the binary while sneaking around 
the secondary black hole. While a part of the gas accretes at around the periastron
onto the primary black hole moving around the center of mass of the binary system,
another part of the gas accretes onto the secondary black hole at the periastron passage.
The resultant superposed light curve exhibit one double-peaked flare per binary orbit,
and its interval between peaks is shorter than those in Models A2, A3, and A4.

%
%%%%%%%%%%%%%%%%%%
\subsection{Accretion flow patterns}
%%%%%%%%%%%%%%%%%%
%
%%%%%%%%%
%% Figure 3  %%
%%%%%%%%%
%
\begin{figure}
\resizebox{\hsize}{!}{
\includegraphics[]{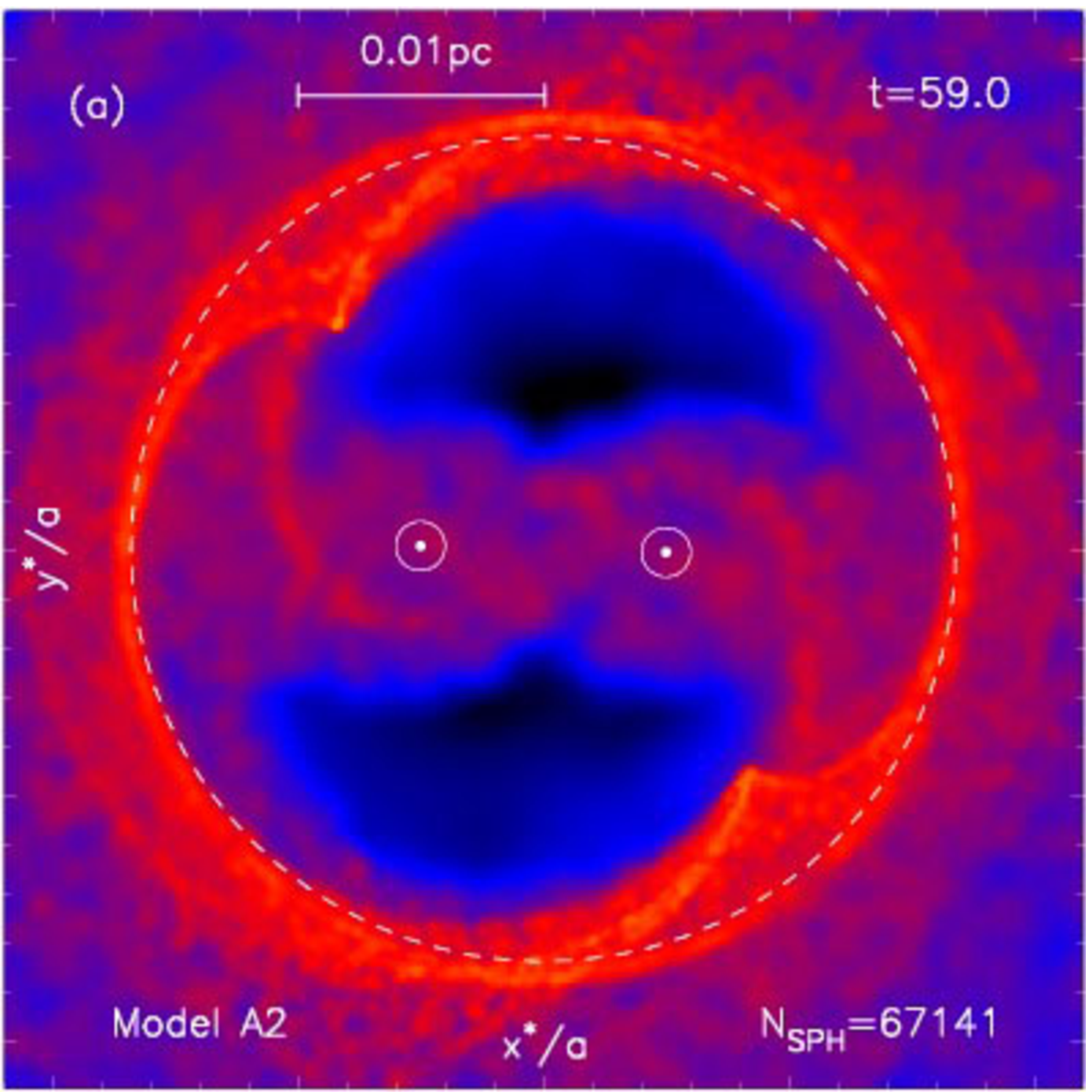}
\includegraphics[]{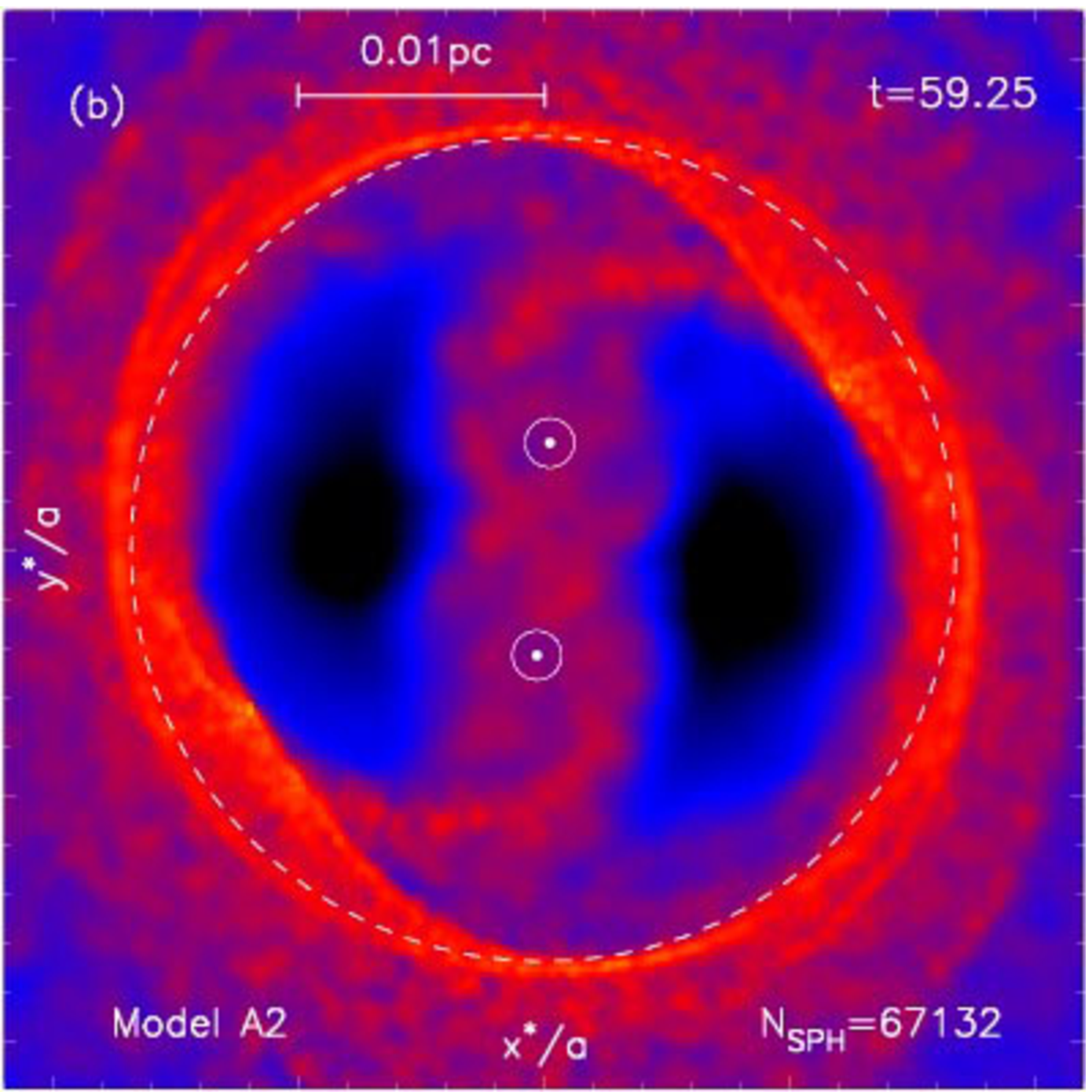}
\includegraphics[]{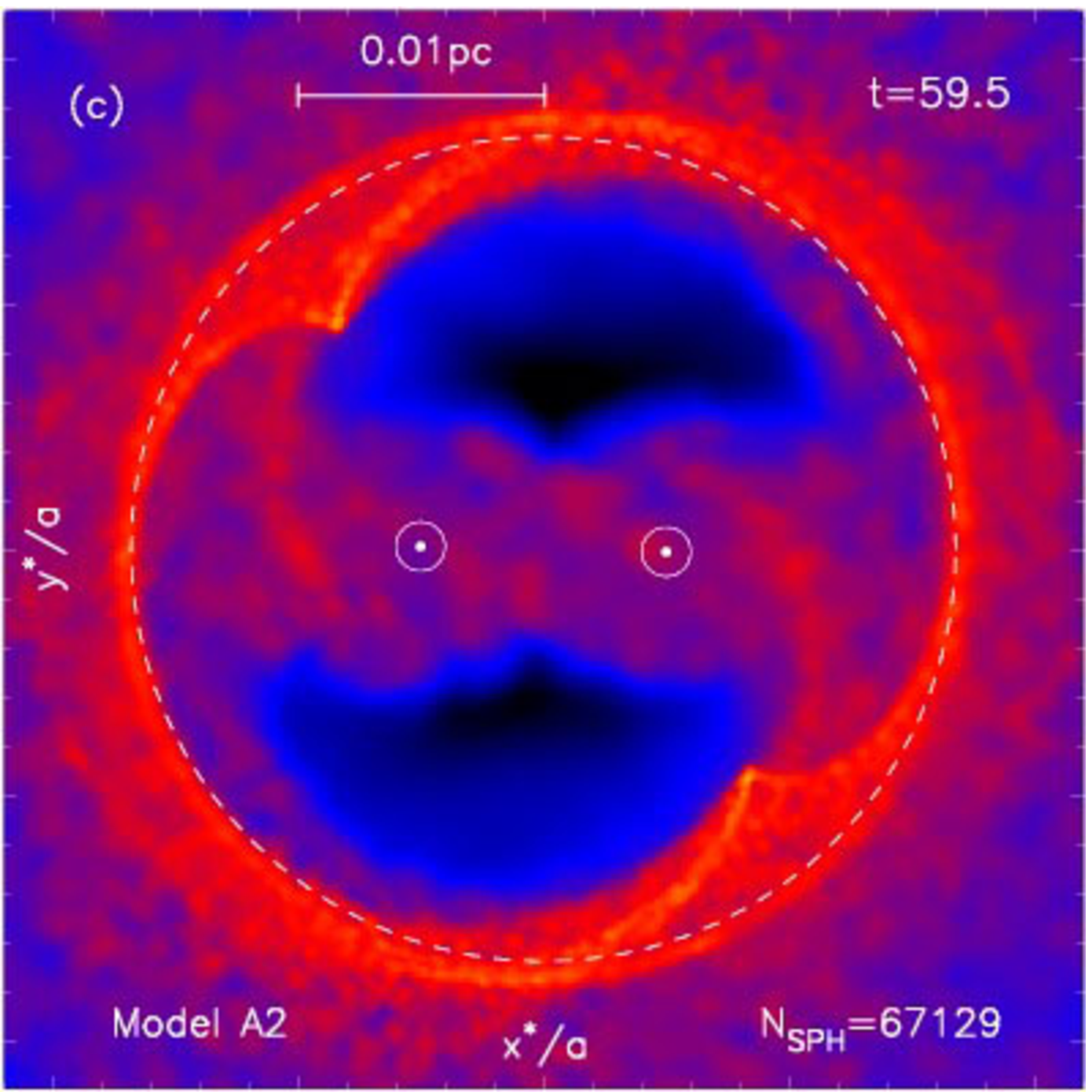}
\includegraphics[]{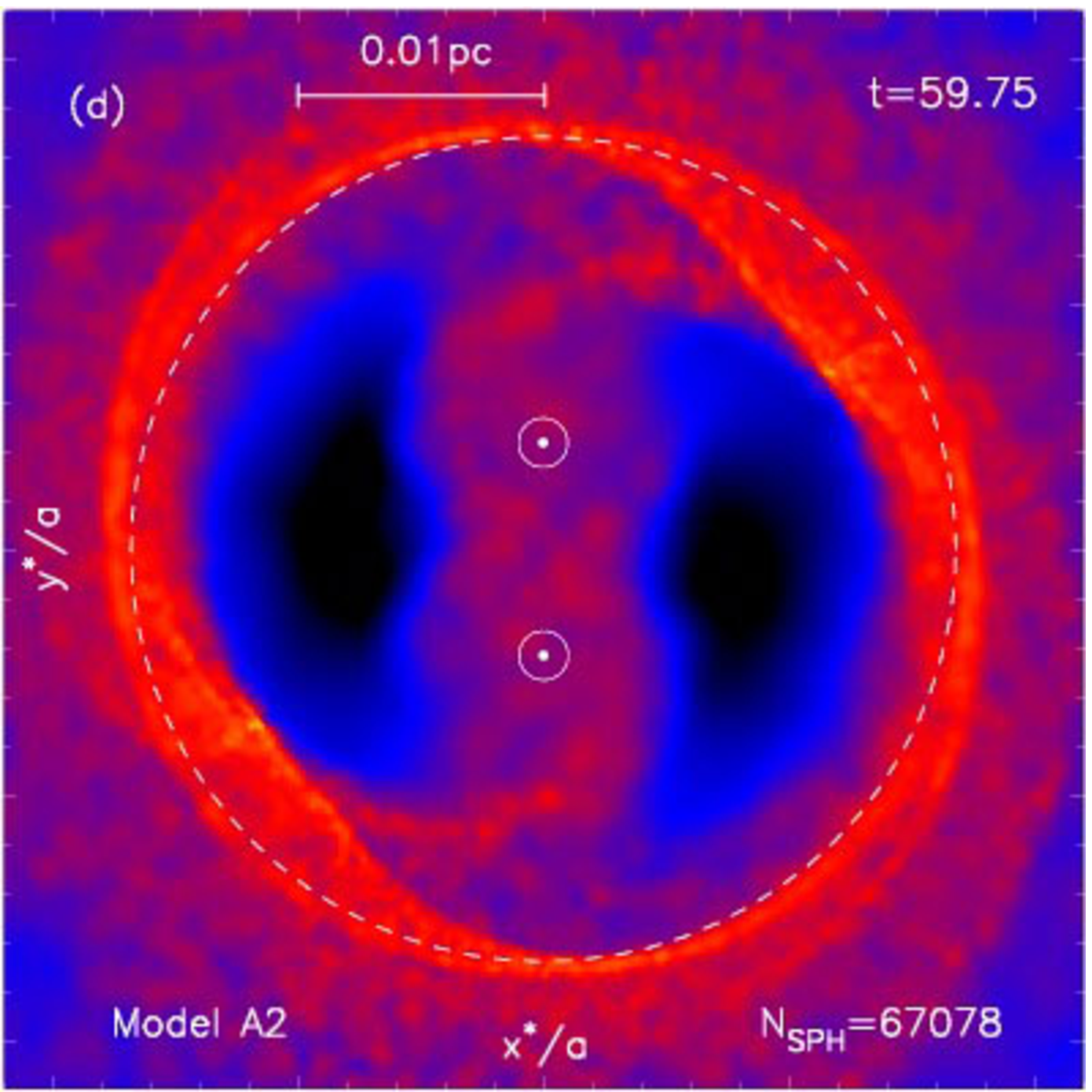}
}\\
\resizebox{\hsize}{!}{
\includegraphics[]{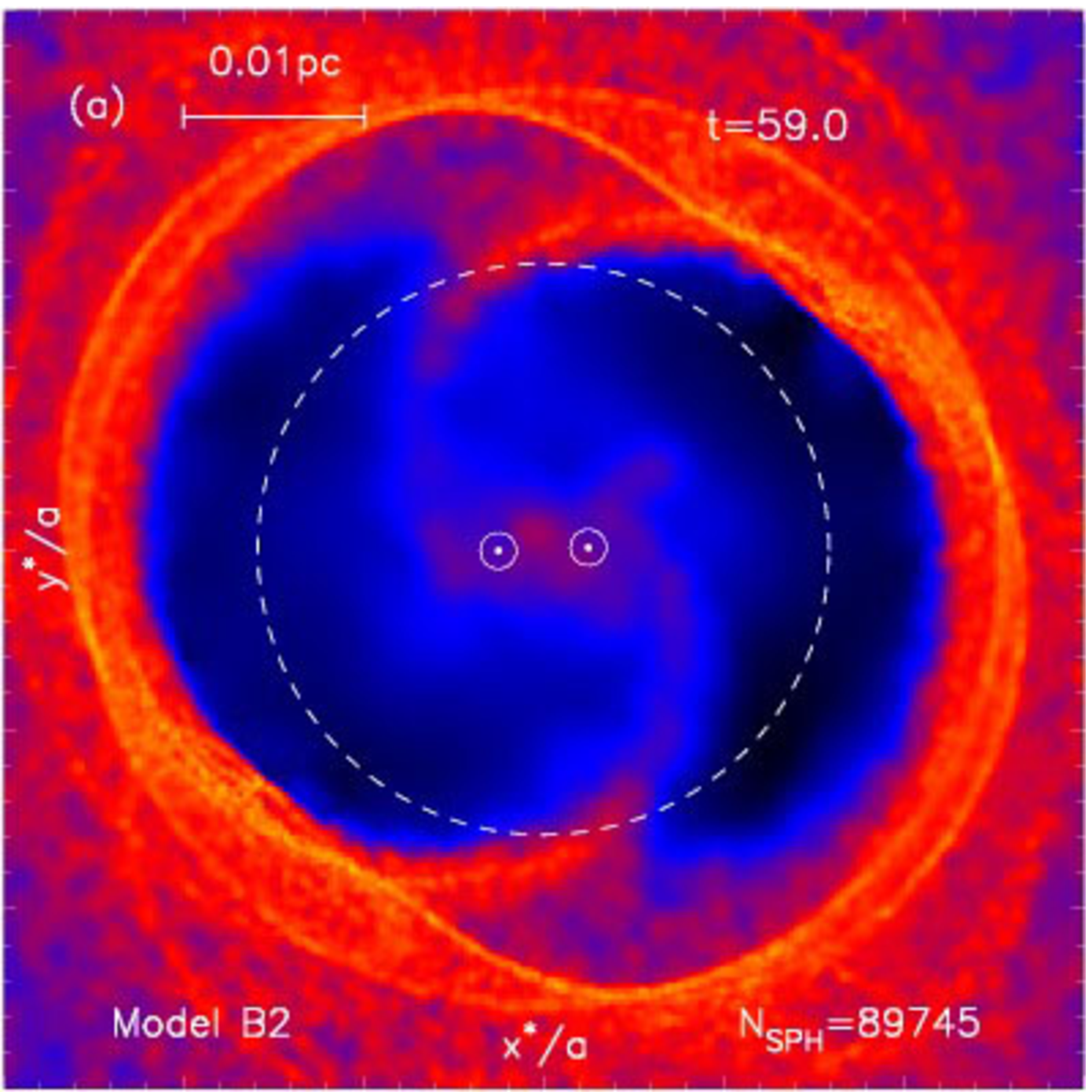}
\includegraphics[]{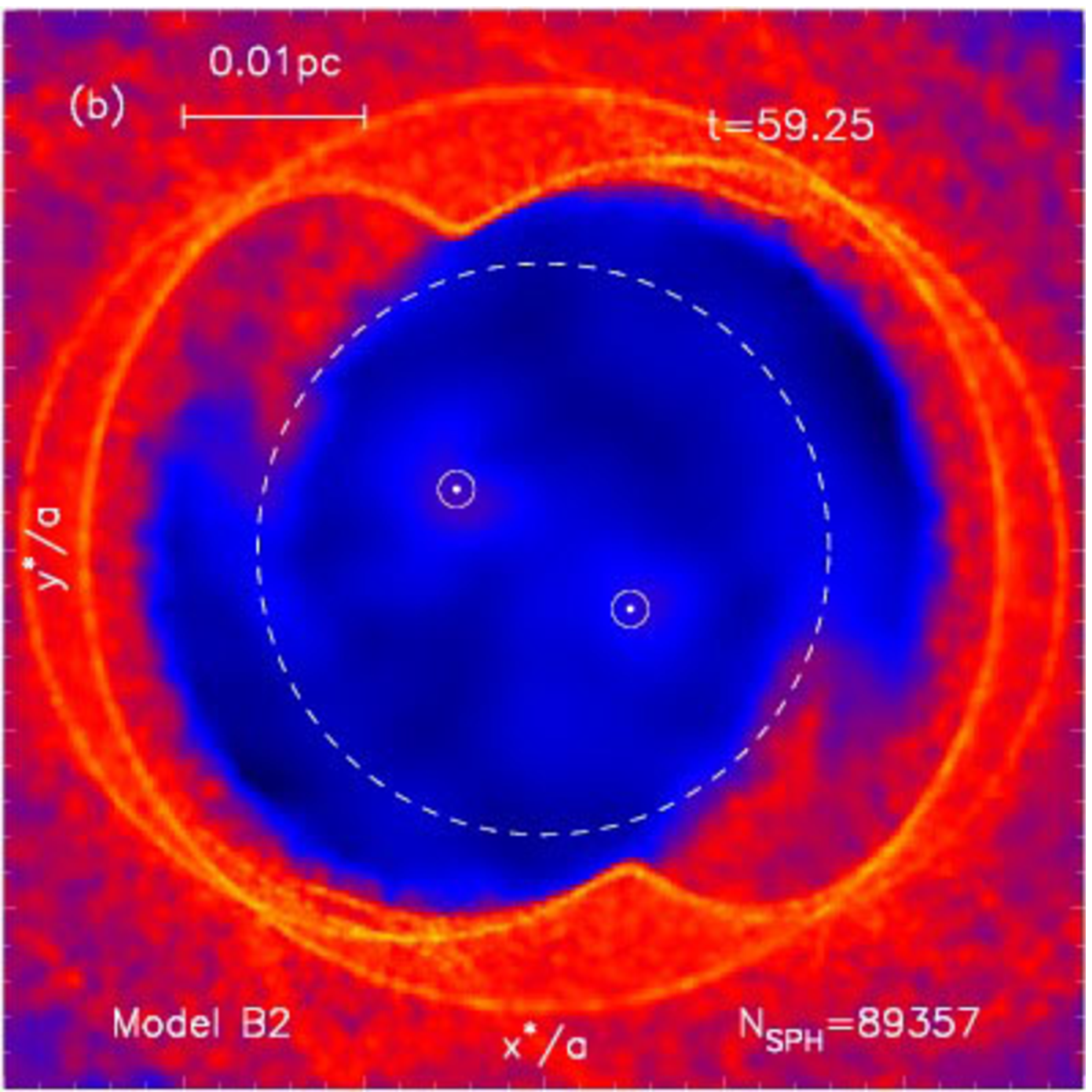}
\includegraphics[]{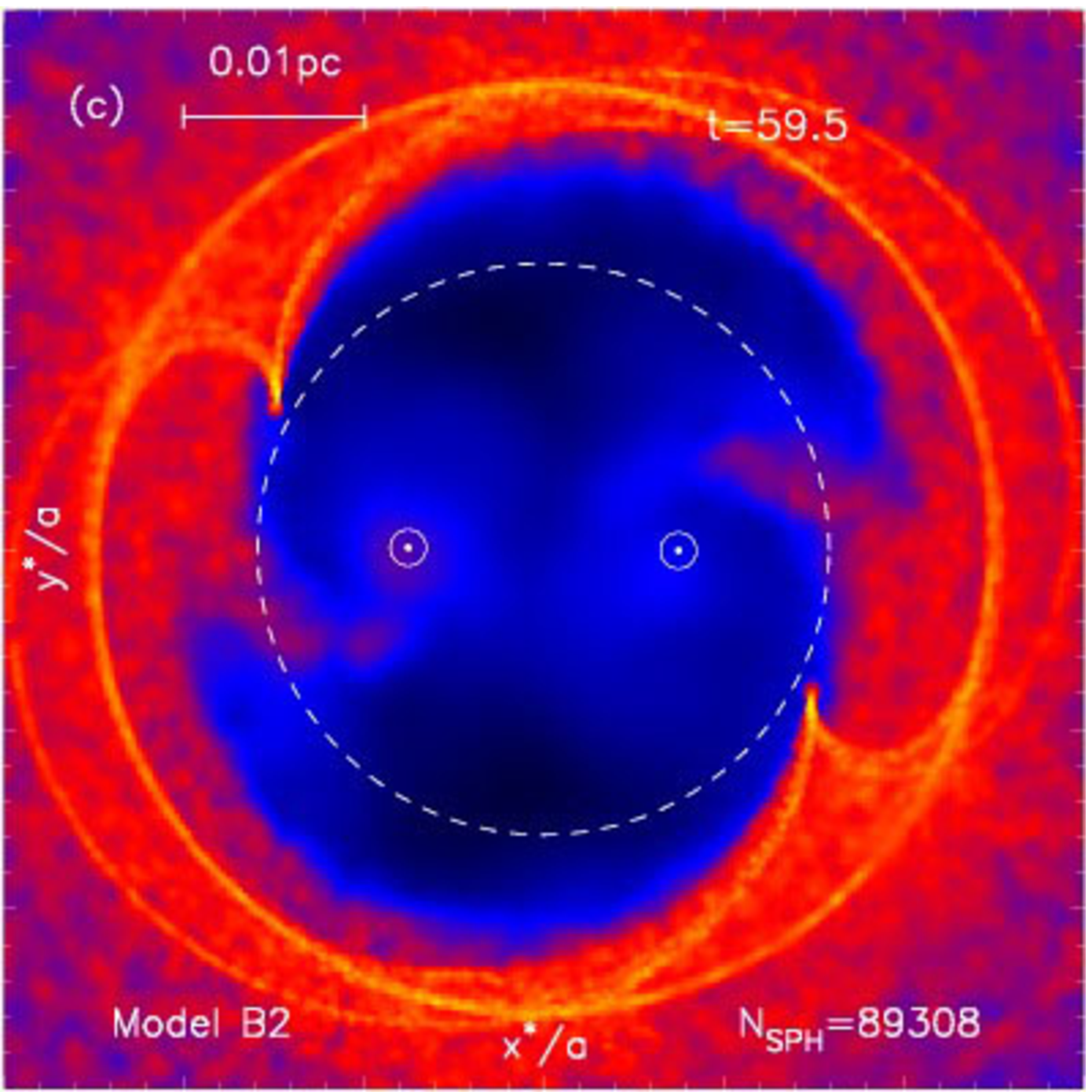}
\includegraphics[]{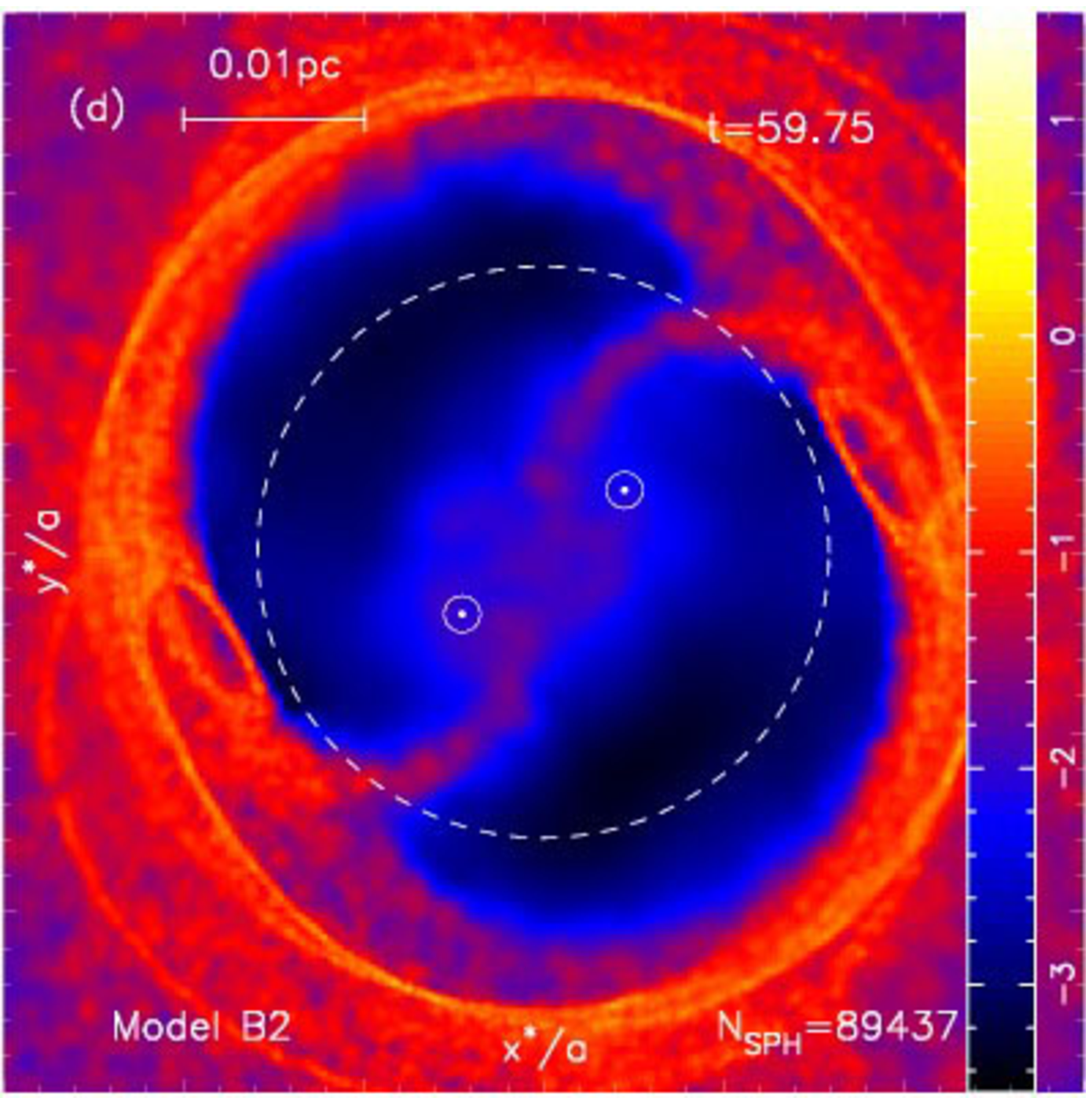}
}
\caption{
A sequence of snapshots of the accretion flow from a circumbinary disk onto binary SMBHs in Model A2 (upper four panels) and Model B2 (lower four panels). The color contours of surface density are displayed in the chronological order from the left (a) to the right (d) for each model. The surface density is calculated by integrating density in the direction of the circumbinary disk plane. The two small points and solid circles denote the positions of two black holes and the accretion radii surrounding them, respectively. The figure is shown in the inertial frame; i.e., both of the binary and the circumbinary disk are rotating around the center in the anti-clockwise direction. 
Note that the disk plane is tilted by $\beta=\pi/6$ from the binary orbital plane with $\gamma = \pi/2$ in both models. The density levels of each panel can be seen in the color chart ($-3.5\le\log\Sigma\le1.5$) at the right side of lower panel (d). The dashed circle represents the initial inner-edge radius $r_{\rm{in}}=1.68a$ for Molde A2, while it is the 1:3 outer Lindbrad resonance radius $\sim2.1a$ for Moldel B2. Annotated in each panel are the major scale in units of $0.01\rm{pc}$, time in units of $P_{\rm orb}$ and the number of SPH particles $N_{\rm SPH}$, respectively.
}
\label{fig:modelA2B2}
\end{figure}
%%%%%%
%

%
% 1st paragraph
In order to visualize how gas accretion onto binary SMBHs occurs, 
we show in Figure 3 sequences of surface density contours for Model A2 (upper panels) 
and Model B2 (lower panels) at four different phases; at the phases of $t=59.0$, $59.25$, 
$59.5$, and $59.75$ from the left to the right, respectively. These panels are illustrated 
in the inertial frame, and both of the black holes and the circumbinary disk are rotating 
in the counterclockwise direction. Note that the black holes are rotating more rapidly than 
the gas in the circumbinary disk according to the Kepler's third law. The two small points 
represent the locations of the primary black hole and secondary black hole, respectively.
The solid small circles surrounding them represent their accretion radii, which are set to be 
$r_{\rm acc}=0.1a$ from the center of each black hole. The dashed circles in both models 
represent the inner edge of circumbinary disk $\sim1.68a$ for Model A2 and the 1:3 outer 
Lindblad resonance radius $\sim2.1a$ for Model B2 (see \cite{al94}), respectively. These 
density maps are projected to the circumbinary disk plane (denoted as the $x^{*}$-$y^{*}$ plane).

%
% 2nd paragraph
First, we decribe the case of Model A2 (upper panels). Although the flow patterns shown 
in the upper four panels look quite similar to those in Model A1 reported by \citet{hms07}, 
we notice some differences between them. At the elapsed time of $t=59.0$ and $59.5$ as 
shown in panels (a) and (c), the innermost part of the circumbinary disk is most strongly 
distorted by the tidal force of each black hole, and therefore the surface density of the 
innermost part of the circumbinary disk is enhanced at the two positions of $\sim 3\pi/4$ 
and $\sim 7\pi/4$ from the $x^*$ axis, where cusp structures are observed.
This is because each black hole is located within the circumbinary disk plane at those times.
By contrast, at the elapsed time of $t=59.25$ and $59.75$ as shown in panels (b) and (d), 
the cusp density structure disappears, though density enhancements are seen to some extent.
This is because each black hole is most distant from the circumbinary disk plane. 
Now we understand that the time variations of the overall density distribution repeat every 
half orbital period, which is the period of the passage of the black holes across the circumbinary 
disk plane. This is responsible for the double-peaked variations in the mass accretion rates.

% final paragraph
Let us, next, examine the case of Model B2, which is illustrated in the lower four 
panels in Figure \ref{fig:modelA2B2}. At the phase of $t=59.25$ as shown in panel (b), 
the separation between the two black holes is going to increase and, hence, 
black holes are approaching the inner edge of the circumbinary disk. At $t=59.5$ 
the binary separation reaches its maximum and the distance to the circumbinary disk 
is at minimum (see panel (c)). The gas in the innermost part of the circumbinary disk 
is pulled out by the black hole from $t=59.25$ to $59.5$, thereby a tidal tail being formed 
and extending inward from the innermost part of the circumbinary disk. This tidal tail 
continuously grows, and at the phase of $t=59.75$ as shown in panel (d) we see a bridge 
connecting the two tidal tails extending from the opposite sides of the circumbinary disk. 
Gas is supplied to the black hole in a next moment, which is clearly seen in panel (a). 
In fact, it takes about a half of binary period for gas to fall onto the black holes from the 
inner edge of the circumbinary disk. This is the reason why mass accretion rate reaches 
its maximum value at around phase zero. It should be noted that such a gas dynamical 
behavior is very similar to that in Model B1, which was reported previously \citep{hms07}. 
In other words, the effects of the tilt angle $\beta$ is not so appreciable for high eccentricity 
cases. We will discuss why it is so in section~\ref{sec:sam}.

%
%%%%%%%%%%%%%%%%%%%%%%%%%%%%%%%%%%
\subsection{Averaged mass accretion rates and circularization radii}
\label{subsec:ave}
%%%%%%%%%%%%%%%%%%%%%%%%%%%%%%%%%%
%

% 1st paragraph
In Table~\ref{tb:t3}, we summarize the averaged mass-accretion rates 
and averaged circularization radii around black holes, and the number 
of SPH particle at the end of run for all models, where the circularization 
radius, $r_{\rm{c},{\it i}}$, is defined by $r_{\rm{c},{\it i}}=j_{i}^2/GM_{\it i},
\hspace{2mm}[{\it i}=1,2]$ and $j_{\it i}$ represents the specific angular 
momentum of the SPH particles which enter inside each accretion radius. 
Thus, their specific angular momentum is proportional to $1/2$ power of 
circularization radius.

% 2nd paragraph
It is noted from Table~\ref{tb:t3} that the same amount of mass injection is 
assumed for Models A1--A4, but the nevertheless the averaged accretion 
rate is higher in Model A1 than others. These trends are also seen in Models B and C.

% final paragraph
The averaged circularization radii are about two orders of magnitude 
larger than the Schwarzshild radius corresponding to each black hole 
mass. This suggests that the averaged circularization radius of infalling 
material indicates the size of an accretion disk formed around each black hole. 
We also note that the averaged circularization radii for the case of $\beta\neq0$ 
is larger than those for the case of $\beta=0$ in all models. This is because 
the distance between each black hole and its nearest inner edge of the 
circumbinary disk is longer in the misaligned system ($\beta\neq0$) than 
that of the coplanar system ($\beta=0$). 
%In Models~C, the circularization radius around the secondary (less massive) black hole is smaller than that of the primary (massive) one.  
The mass tidally stripped from the circumbinary disk does not directly 
accrete onto each black hole but via the accretion disk around each black hole. 
We will briefly discuss how the accretion disk evolve in section~\ref{sec:dis}.

%
%%%%%%%%%%%%%%%%%%%%
\section{Simple semi-analytical models}
\label{sec:sam}
%%%%%%%%%%%%%%%%%%%%

%
%%%%%%%
% Figure 4
%%%%%%%
%
\begin{figure}
\begin{center}
\includegraphics[width=12cm]{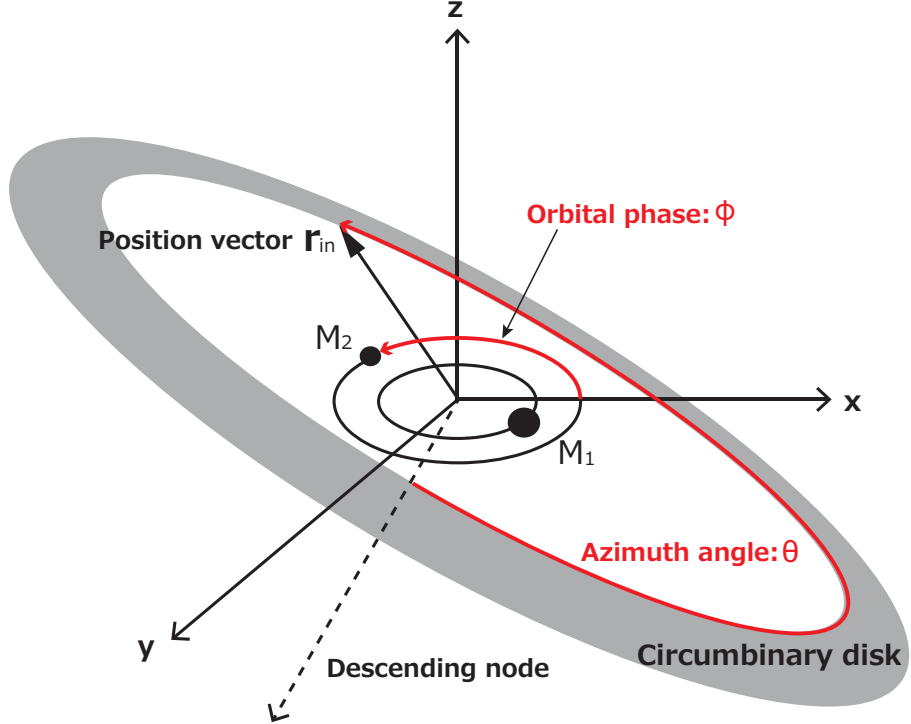}
\end{center}
\caption{
Schematic view explaining the definitions of the two angles ($\phi, \theta$) 
and the descending node. The orbital phase $\phi$ and the azimuth angle 
$\theta$ are measured from $x$-axis and the descending node, respectively.
}
\label{fig:pv_angles}
\end{figure}

% 1st paragraph
To understand in a simpler way how a variety of variation patterns 
in mass accretion rates arises, we construct a semi-analytical model. 
Since mass accretion occurs by tidal stripping of gas from the inner 
edge of the circumbinary disk, and since the gravitational attraction 
force to the circumbinary disk is strongest when the distance between 
the black hole and its nearest inner edge of the circumbinary disk is at 
a minimum, we can guess the number of peaks per orbital period by 
calculating the minimum distance as a function of the orbital phase 
for parameter sets of all models.

% 2nd paragraph
The position vector of the inner edge of the circumbinary disk, 
$\mbox{\boldmath r}_{\rm{in}}$, can be expressed by
\begin{equation}
\mbox{\boldmath $r$}_{\rm{in}}=(r_{\rm{in}}[\cos\theta\sin\gamma+\sin\theta\cos\gamma\cos\beta], 
r_{\rm{in}}[\sin\theta\sin\gamma\cos\beta-\cos\theta\cos\gamma], -r_{\rm{in}}\sin\theta\sin\beta),
\label{eq:rin}
\end{equation}
where the orbital phase $\phi$ and the azimuth angle $\theta$ 
are measured from $x$-axis and the descending node, respectively (see Figure~\ref{fig:pv_angles}).
From equations~(\ref{eq:ri}) and (\ref{eq:rin}), we obtain the formula 
for the distance: 
\begin{eqnarray}
d(\beta,\gamma,e,q; \phi)=|\mbox{\boldmath $r$}_{\rm{in}}-\mbox{\boldmath $r$}_{i}|
=\sqrt{4a^2 + r_{i}^2-4ar_{i}[\sin\theta\cos(\phi-\gamma)\cos\beta-\cos\theta\sin(\phi-\gamma)]},
\label{eq:d}
\end{eqnarray}
where the value of $\theta$ for a given $\phi$ is numerically chosen so as to 
give the minimum distance. Here, we assign $i=2$, since the less massive 
black hole can move on a larger extent, thereby getting closer to the circumbinary 
disk than the massive one. 

\begin{figure}
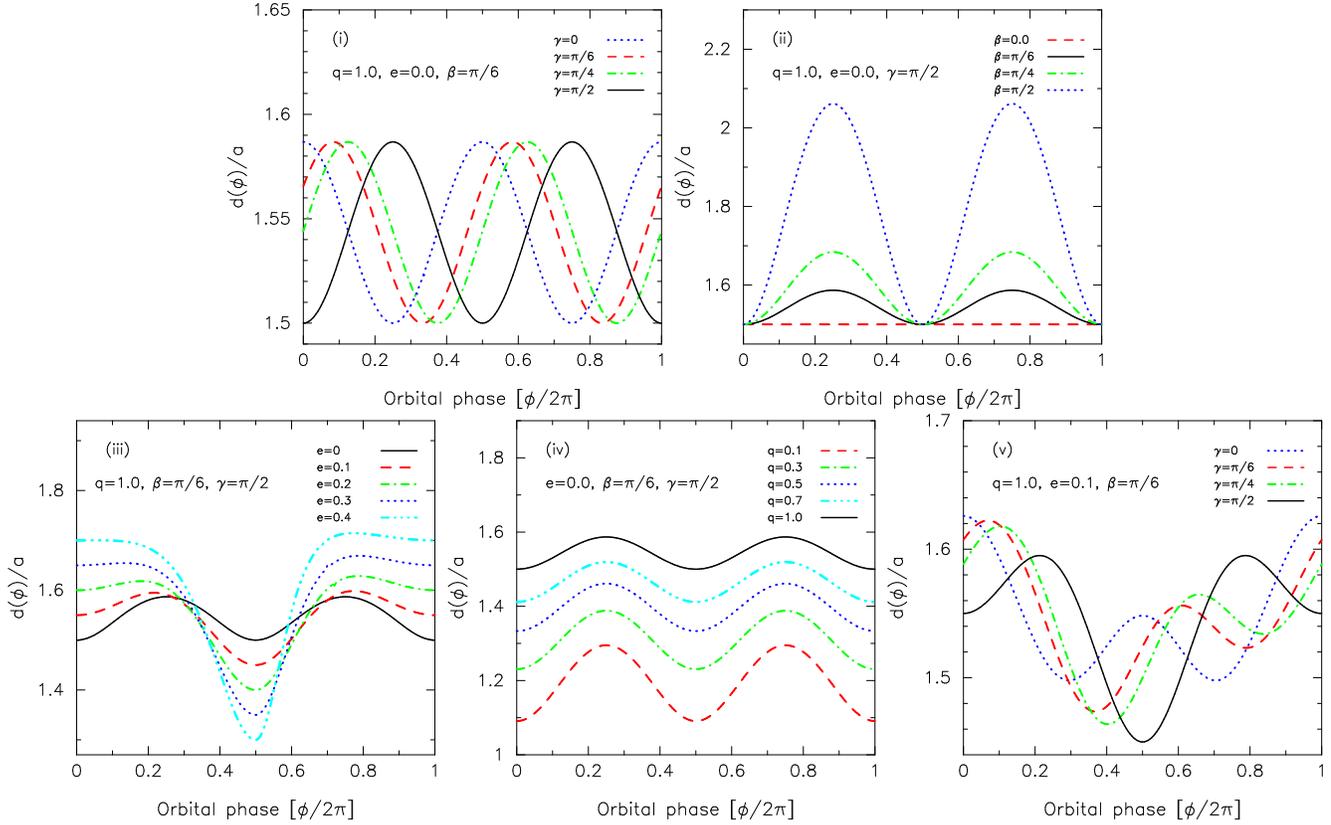

\begin{center}
\includegraphics[scale=0.50]{figure5a.ps}
\includegraphics[scale=0.50]{figure5b.ps}\\
\includegraphics[scale=0.50]{figure5c.ps}
\includegraphics[scale=0.50]{figure5d.ps}
\includegraphics[scale=0.50]{figure5e.ps}
\end{center}
\caption{
Orbital phase dependence of the normalized distance, 
$d(\beta,\gamma,e,q; \phi)/a$, between 
the black hole and the inner edge of circumbinary disk.
The dependence of the normalized distance on the four parameters: 
$\beta$, $\gamma$, $e$, and $q$ are shown in panels (i)--(v).
In all the panels, the solid (black) line indicates the fiducial model; 
$(\beta, \gamma, e, q)=(\pi/6, \pi/2, 0.0, 1.0)$ for panels (i)--(iv) and 
$(\beta, \gamma, e, q)=(\pi/6, \pi/2, 0.1, 1.0)$ for panel (v). 
For the meanings of other lines, see text and the explanations in each panel.
}
\label{fig:sep}
\end{figure}

% 3rd paragraph
Figure~\ref{fig:sep} shows the orbital phase dependence of the distance between 
the black hole and its nearest inner edge of circumbinary disk.
There are four different parameters: orbital eccentricity ($e$), mass ratio ($q$), 
tilt angle ($\beta$), and azimuth of tilt ($\gamma$).
The dependence of $d(\beta,\gamma,e,q; \phi)$ normalized by the semi-major axis 
on these parameters are shown in panels (i)-(v).
The fiducial parameters are those of Model~A2; i.e., $(\beta,\gamma,e,q)=(\pi/6,\pi/2,0.0,1.0)$.

% 4th paragraph
Panel (i) shows $\gamma$-dependence of normalized distance, $d(\beta,\gamma,e,q; \phi)/a$, 
for fixed values of $(q,e,\beta)=(1.0,0.0,\pi/6)$. The solid line, dotted line, dashed line, and 
dash-dotted line represent $d(\phi)/a$ for $\gamma=\pi/2$, $0$, $\pi/6$, and $\pi/4$, respectively. 
We note that $d(\phi)/a$ is the shortest twice per binary orbit and that the phases of the minimum 
distance shift by varying the parameter $\gamma$. Here, let us go back to see the difference 
between mass accretion rate of Model~A2 and that of Model A4. The phase difference between 
the solid line ($\gamma=\pi/2$) and the dotted line ($\gamma=0$) is $\pi/2$. This supports our 
simulation results of Models A2 and A4.

% 5th paragraph
Panel (ii) shows $\beta$-dependence of the normalized distance for the fixed values of 
$(q,e,\gamma)=(1.0,0.0,\pi/2)$. The solid line, dashed line, dash-dotted line, and dotted 
line represent $d(\phi)/a$ for $\beta=\pi/6$, $0$, $\pi/4$, and $\pi/2$, respectively.
The normalized distance reaches its minimum value twice per binary orbital period 
except for the case with $\beta=0$, in which $d(\phi)/a$ is constant. That is, the non-zero 
values of $\beta$ is essential to produce two peaks per binary orbit in the mass-accretion-rate 
variations. This panel provides a reasonable explanation about why the results of Models 
A2-A4 are shown in the left panel of Figure~\ref{fig:mdotL}.

% 6th paragraph
Panel (iii) shows $e$-dependence of the normalized distance for the fixed values of 
$(q,\beta,\gamma)=(1.0,\pi/6,\pi/2)$. The solid line, dashed line, dash-dotted line, 
dotted line, and three-dotted line represent $d(\phi)/a$ for $e=0$, $0.1$, $0.2$, $0.3$, 
and $0.4$, respectively. The normalized distance has two round peaks for $e=0$, but
the larger $e$ is, the deeper becomes a hollow at around $\phi =\pi$.
Although we have mentioned that non-zero values of $\beta$ give rise to the double 
peaked modulations in mass accretion rates, this is not always the case when the 
orbital eccentricity is not zero. This will explain that the mass accretion rates of 
Models~B2, B3, and B4 have a single peaked shape. We have also calculated the 
cases with $\gamma=0.0$, reaching the same conclusion; i.e., single peaked variations 
are found for $e \gtsim 0.3$. This supports the result of Model~B4.

% 7th paragraph
Panel (iv) shows $q$-dependence of the normalized distance between the 
secondary (less massive) black hole and its nearest inner edge of the 
circumbinary disk for the fixed values of $(e,\beta,\gamma)=(0,\pi/6,\pi/2)$. 
The solid line, dashed line, dash-dotted line, dotted line, and three-dotted 
line represent $d(\phi)/a$ for $q=1.0$, $0.1$, $0.3$, $0.5$, and $0.7$, 
respectively. The lower the binary mass ratio is, the shorter becomes $d(\phi)/a$. 
Variations in the mass ratio do not change the number of peaks of the normalized 
distance. Obviously, the smaller $q$ is, the smaller becomes the mean distance. 
This is because the semi-major axis of the secondary black hole 
($r_2$ in equation~\ref{eq:rphi})  increases as $q$ decreases.

% final paragraph
Panel~(v) shows the same $\gamma$-dependence as those in panel~(i) 
but for $e=0.1$. This panel shows how the normalized distance changes 
with $\gamma$ for the cases with an even smaller orbital eccentricity. 
The solid line, dotted line, dashed line, and dash-dotted line represent 
$d(\phi)/a$ for $\gamma=\pi/2$, $0$, $\pi/6$, and $\pi/4$, respectively. 
The solid line is the same as the dashed line of panel (iii).
It is interesting to note that double peaked variation curve for the case of 
$\gamma=0$ changes to a single peaked one as $\gamma$ increases.

%
%%%%%%%
%%%%%%
%
%
%%%%%%%%%%
\section{Discussion}
\label{sec:dis}
%%%%%%%%%%
%
% 1st paragraph
Periodic light variations are not generally observed from single SMBH systems. 
There is a report for the presence of quasi-periodic oscillations (QPOs) in AGNs 
\citep{gie08}, but the QPOs do not mean periodic occurrence of clear-cut flares 
but mean rather gradual variations. Unlike single SMBH systems, periodic light 
variations can be seen in an eccentric binary black hole system because of 
periodically enhanced binary-disk interactions. In most of previous studies, 
it was implicitly assumed that the binary orbital plane is aligned with the 
circumbinary disk plane. Since there are no strong reasons to believe the 
binary-disk alignment, it is natural to relax this assumption and to examine 
what light variations are expected for the misaligned system. As shown in 
Figure~\ref{fig:mdotL}, a circular binary surrounded by a misaligned circumbinary 
disk exhibits a double peak structure in the variations of mass accretion rates, 
whereas an eccentric binary with a moderately large orbital eccentricity shows 
a single peak per orbit, even if the circumbinary disk is inclined from the binary orbital plane.

% 2nd paragraph 
%%%%%%%%%%%%%%%%%%%%%%%%%%%%%%%%%%%%
\subsection{Inner edge radius of misaligned circumbinary disks}
%%%%%%%%%%%%%%%%%%%%%%%%%%%%%%%%%%%%
%In contrast to the retrograde circumbinary disk case, 
The radius of the inner edge of the circumbinary disk is determined by the balance 
between the viscous torque of the prograde circumbinary disk and the tidal/resonant 
torque which acts on it \citep{al94}. In the case of moderate orbital eccentricity, the 
typical inner edge radius is at the $1:3$ outer Lindblad resonance radius, which is 
estimated to be $\sim2.1a$. If the circumbinary disk is misaligned with respect to the 
binary plane, the resonant torque will be weaker than otherwise, which makes the 
inner edge smaller. In order to more precisely determine the size of the inner edge 
of the circumbinary disk, we need to investigate how the tidal/resonant torque acts 
on the misaligned circumbinary disk. This is also essential to calculate the circumbinary 
disk structure.

% 4th paragraph
%%%%%%%%%%%%%%%%%%%%%%%%
\subsection{Triple disk model for OJ 287}
%%%%%%%%%%%%%%%%%%%%%%%%

A blazer OJ287 exhibits quasi-periodic optical outbursts of 12 year interval.
In order to explain such quasi-periodic nature, a binary black hole model 
was firstly proposed by \citep{sill88}. \citet{lv96} subsequently proposed the 
modified model that a less massive black hole orbits around a more massive 
one with undergoing relativistic precession and impacts twice per binary orbit 
on the accretion disk around the more massive black hole. This model can 
explain the quasi-periodicity of outbursts with a double peak structure at a few 
year interval.

% 5th paragraph
As shown in Models C1 and C2 at the right panel of Figure~\ref{fig:mdotL}, 
the solid line and dashed (red) line represent the mass accretion rates of 
the primary black hole and that of the secondary black hole, respectively. 
Respective mass accretion rates periodically vary with time and have a single 
peak around the periastron but slightly different orbital phases.The resultant 
combined mass accretion rate provides a periodic double peaked structure 
with a short interval. We, therefore, propose an alternative binary black hole 
model that the circumbinary disk around an eccentric binary composing of 
black holes with different masses is misaligned by a relatively small tilt angle ($0\lesssim\beta\lesssim\pi/6$).

% 6th paragraph
%%%%%%%%%%%%%%%%%%%%%%%%%%%%%%%%%%%%%%%%
\subsection{Precessions of misaligned circumbinary disks}
%%%%%%%%%%%%%%%%%%%%%%%%%%%%%%%%%%%%%%%%

The binary-disk interaction also gives rise to a precession of the circumbinary disk 
\citep{mm08,khato09,dza12}. With the assumption that $r\gg a$ and $M_2\ll{M_1}$, 
the precession frequency of the misaligned circumbinary disk is given by (e.g., \cite{nix11b}) as
\begin{eqnarray}
\frac{P_{\rm{prec}}}{P_{\rm{orb}}}\approx\frac{4}{3}\frac{1+q}{q}\left(\frac{r}{a}\right)^{7/2}\frac{1}{|\cos\beta|}\hspace{2mm}(-\pi/2<\beta<\pi/2),
\label{eq:prec}
\end{eqnarray}
where $P_{\rm{prec}}$ and $P_{\rm{orb}}$ are the precession period and binary 
orbital period, respectively. It is clear that $P_{\rm{prec}}$ is much longer than 
$P_{\rm{orb}}$, e.g. $P_{\rm{prec}}\gtrsim100P_{\rm{orb}}$ for $q=0.1$ and $r=2a$.
%While the binary potential varies with the orbital phase, the circumbinary disk slowly precesses \citep{mm08,khato09,dza12}.
This can produce a light variation in circumbinary disk with a beat period 
$P_{\rm{beat}}$, where it is expressed by $1/P_{\rm{beat}}\equiv1/P_{\rm{orb}}-1/P_{\rm{prec}}$. 
From equation~(\ref{eq:prec}), we obtain the beat period normalized by the binary orbital period:
\begin{equation}
\frac{P_{\rm{beat}}}{P_{\rm{orb}}}=\left[1-\frac{3}{4}\left(\frac{r}{a}\right)^{-7/2}\frac{q}{1+q}|\cos\beta|\right]^{-1}
\label{eq:beatp}
\end{equation}
which is slightly longer than the orbital period, as far as $r\gg{a}$. 
The beat period for the case of $q=0.1$ and $r=2a$ is, for example, approximately $1.01P_{\rm{orb}}$. 
In a subsequent paper, we will study this topic in more detail.

% final paragraph
%%%%%%%%%%%%%%%%%%%%%%%%%%%%%%%%%%%%%%%%
\subsection{Luminosity variations}
%%%%%%%%%%%%%%%%%%%%%%%%%%%%%%%%%%%%%%%%

Another interesting topic is how the mass finally accretes onto each black hole via 
an accretion disk from the circumbinary disk. In section~\ref{subsec:ave}, we show 
that the infalling material is circularized around each black hole (the estimated 
circularization radii can be seen in Table~\ref{tb:t3}). This leads to the formation of 
the triple disk system, which is composed of two accretion disks around black holes 
and one circumbinary disk surrounding them \citep{hmh08}. However, it is poorly 
known what structure each accretion disk has and how it evolves. Once these two 
accretion disks are formed by gas supply from the circumbinary disk, they viscously 
evolve and gas in the accretion disk finally accretes onto each black hole after the 
viscous timescale. Assuming that the accretion disk is the standard disk, the viscous 
timescale measured at the circularization radius is much longer than the binary 
orbital period. The precise shape of light curves may be different from that of variations 
in mass accretion rates. It is interesting to examine the basic properties of radiations emitted 
from each accretion disk.

%
%%%%%%%%%%%
\section{Conclusions}
%%%%%%%%%%%
%

We have carried out numerical simulations of accretion flows 
from a circumbinary disk which is inclined from the binary orbital plane, 
in order to examine to what extent the basic properties of mass-accretion-rate 
variations may alter, compared with the coplanar cases. Our main conclusions are 
summarized as follows:
\begin{enumerate}
\renewcommand{\theenumi}{\arabic{enumi}}
\item 
We find that the mass accretion rates exhibit a double peak per binary orbit 
in a circular binary system, when the circumbinary disk is misaligned with the binary 
orbital plane. This is because each black hole passes across the circumbinary disk 
plane twice per binary orbit and then attracts the gas there. 
This double peak nature of mass accretion rates is also independent of the azimuth of tilt. 
The tilt angle is one of important orbital parameters to determine variation patterns of radiations 
emitted from a binary black hole system.
\item 
The orbital eccentricity remains to be an important orbital parameter to produce 
single sharply peaked variations per binary orbit in mass accretion rates even in a misaligned circumbinary disk system. This is because each black hole is closest to the inner edge of circumbinary disk 
once per binary orbit in most cases. The simple semi-analytic model (see section~\ref{sec:sam}) predicts that 
this single peak nature is independent of both the tile angle and the azimuth of tilt, as long as $e\gtrsim0.3$.
\item
In the case of an eccentric binary composing of black holes with different masses,
the less massive black hole can get closer to the circumbinary disk than the massive one, 
thus tidally splitting gas from its inner edge, but the created gas flows are comparably 
captured by both black holes with a short time delay. The superposed accretion rates 
show periodic outbursts with an apparent double peaked structure with a short interval.
\end{enumerate}

%%%%
\bigskip
%%%%

KH thanks Atsuo~T. Okazaki for helpful suggestions. 
The numerical simulations reported here were 
performed using the computer facilities of Yukawa 
Institute of Theoretical Physics, Kyoto University. 
This work is supported in part by the Grants-in-Aid of the Ministry of 
Education, Science, Culture, and Sport and Technology (MEXT) 
[21540304, 22340045, 22540243, 23540271 KH, 22340045 SM], 
and the global COE programs on The Next Generation of Physics, 
Spun from Diversity and Emergence from MEXT.

%%%%%%%%%%%%

\end{document}